\documentclass[aps,prl,twocolumn,superscriptaddress,groupedaddress]{revtex4}
\usepackage{graphicx}  % needed for figures
\usepackage{dcolumn}   % needed for some tables
\usepackage{bm}        % for math
\usepackage{amssymb}   % for math

\usepackage{color}

\newcommand{\Lp}{L_{\text{p}}}

\begin{document}

%%%%%%%%%% Merge with supplemental materials %%%%%%%%%%
%%%%%%%%%% Prefix a "S" to all equations, figures, tables and reset the counter %%%%%%%%%%
\setcounter{equation}{0}
\setcounter{figure}{0}
\setcounter{table}{0}
\setcounter{page}{1}
\makeatletter
\renewcommand{\theequation}{\arabic{equation}}
\renewcommand{\thefigure}{\arabic{figure}}
\renewcommand{\thesection}{\Roman{section}}

\widetext
%\leftline{Version of \today}

\title{Non-equilibrium fluctuations of a semi-flexible filament
 driven by active cross-linkers
}

\author{Ines Weber}
\affiliation{Fachrichtung Theoretische Physik, Universit\"at des Saarlandes D-66123 Saarbr\"ucken, Germany}
\affiliation{Laboratoire de Physique Th\'eorique, CNRS (UMR 8627),
Univ. Paris-Sud, Univ. Paris-Saclay,
%B\^atiment 210, F-91405 ORSAY Cedex, France}
91405 Orsay, France}
\author{C\'ecile Appert-Rolland}
\affiliation{Laboratoire de Physique Th\'eorique, CNRS (UMR 8627),
Univ. Paris-Sud, Univ. Paris-Saclay,
%B\^atiment 210, F-91405 ORSAY Cedex, France}
91405 Orsay, France}
\author{Gr\'egory Schehr}
\affiliation{Laboratoire de Physique Th\'eorique et Mod\`eles Statistiques, CNRS (UMR 8626), 
Univ. Paris-Sud, Univ. Paris-Saclay,
%B\^atiment 100, F-91405 ORSAY Cedex, France}
91405 Orsay, France}
\author{Ludger Santen}
\affiliation{Fachrichtung Theoretische Physik, Universit\"at des Saarlandes D-66123 Saarbr\"ucken, Germany}

\date{\today}

\begin{abstract}
The cytoskeleton is an inhomogeneous network of semi-flexible filaments,
which are involved in a wide variety of active biological processes. 
Although the cytoskeletal filaments can be very stiff and embedded in a
dense and cross--linked \ network, it has been shown that, in cells, they 
typically exhibit significant bending on all length scales. In this work we
propose a model of a semi-flexible filament deformed by different types of 
cross-linkers for which one can compute and investigate
the bending spectrum. Our model  allows to couple the
evolution of the deformation of the semi-flexible polymer
with the stochastic dynamics of linkers which exert transversal forces onto the filament.
We observe a $q^{-2}$ dependence of the bending spectrum
for some biologically relevant parameters and
in a certain range of wavenumbers $q$.
However, generically, the spatially localized forcing
and the non-thermal dynamics both introduce
deviations from the thermal-like $q^{-2}$ spectrum.
\end{abstract}

%\pacs{}
\maketitle

% ---------------- INTRO ---------------- INTRO ---------------- INTRO ---------------- INTRO ---------------- INTRO ---------------- INTRO ---------------- INTRO ---------------- INTRO ---------------- INTRO ----------------  %

In recent years many studies were performed on active gels to investigate their complex and dynamic structure, 
which shows generic non-equilibrium behavior. The cytoskeleton, an important example of an active gel, is able 
to form self-organized structures that are the basis of many fundamental processes within cells \cite{alberts2014, bray2000, howard2001}.
The cytoskeleton is composed of actin and intermediate filaments, as well as microtubules, that take important 
roles for example in cell motility, cell division and intracellular transport~\cite{alberts2014, stricker_f_g2010}. It has 
been shown that these cytoskeletal filaments can cross-link via static \cite{selden_p1986} and dynamic interactions \cite{rodriguez2003,straube2006}.

Mechanically the cytoskeletal filaments are semi-flexible filaments with very different persistence lengths. 
 {\it In vitro} measurements  estimated a thermal persistence lengths of the order of $17 \mu$m for actin and
a few millimeters for microtubules \cite{gittes1993,vanmameren2009,brangwynne2007a}.
By contrast, much smaller persistence lengths are observed for microtubules {\em in vivo} ($\approx30\mu m$ 
in~\cite{brangwynne2007b}). These strong deformations are interpreted to be the result of large
non-thermal forces of the order of $1$-$10$pN, which is in the range of  individual motors' strength.
While in some experiments it was surprisingly observed that the bending spectrum exhibits the same shape 
as thermal ones \cite{brangwynne2007b}, other observations reported strong deviations~\cite{brangwynne2008}.

Some continuous theoretical descriptions of active networks exist~\cite{julicher2007,marchetti2013,broedersz_m2014},
which allow to study the deformation of an embedded filament~\cite{kikuchi2009},
However, it is of great interest to understand how deformations originate from
microscopic discrete forcing~\cite{mackintosch_l2008,ronceray_l2015,ronceray_b_l2016,broedersz2010,shekhar2013b,gladrow2016}.

\begin{figure}
\includegraphics[width=\columnwidth]{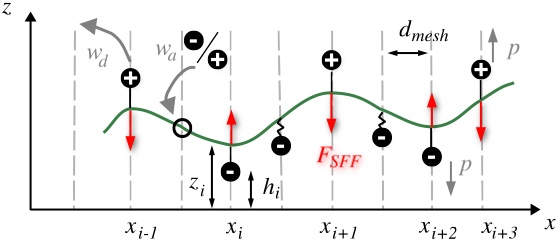}
\caption{\label{fig:MTModel}
Sketch of the model in the case of {\em active} cross-linkers.
Cross-linkers (black dots) are connected to the SFF (green)
through rope-like chains, which can be fully extended (straight segment)
or not (wavy line).
With rate $\omega_{\rm a}$, cross-linkers can attach at empty intersections
($\circ$) between the SFF and the background filaments.
The SFF exerts some forces $F_{\rm{SFF}}$
(red arrows) on fully extended cross-linkers, which
are located at positions $(x_i,h_i)$ and impose the vertical
positions $z_i(x_i)$ of the SFF.
Cross-linkers step upwards or downwards ($(+)$ or $(-)$ labels)
along the polarized vertical background filaments,
with stepping rate $p \equiv p(F_{\rm{SFF}})$.
Cross-linkers detach with rate $\omega_{\rm d} \equiv \omega_{\rm d}(F_{\rm {SFF}})$.
}
\end{figure}

In this paper we consider an idealized system (Fig.~\ref{fig:MTModel})
in which a set of cross-linkers impose transverse
deformations to a semi-flexible filament (SFF).
To couple the dynamics of SFF and linkers,
we determine at each instant
the equilibrium shape of the SFF under the constraints
imposed by the cross-linkers.
The method also provides the forces exerted by the deformed
filament on each cross-linker, allowing to implement
some feedback of the SFF onto the stochastic linkers dynamics.

We shall consider two types of linkers, having thermal or non-thermal
dynamics. This will allow us to disentangle the geometrical effects
from those due to the non-thermal dynamics of active linkers.
We apply our algorithm to explore
the dependence of the persistence length $\Lp$
and of the bending spectrum upon various
parameters, including
the properties of linkers
and of the surrounding network.

We now introduce our model in more details.

{\em Semi-flexible filament} --
The bending energy $E$ of a SFF
of length $L$ with bending rigidity $k$ is given by \cite{landau_l1981a}
\begin{equation}
E=k \int_{0}^{L} \left(\frac{\partial \theta}{\partial s}\right)^2ds
\label{eq:defE}
\end{equation}
and depends on the local curvature ${\partial \theta}/{\partial s}$,
where $\theta(s)$ is the tangent angle and $s$ the contour length. 
We shall express the value of $k$ in units of $k_{MT}$,
the experimentally measured bending rigidity of microtubules (see Table 1 in
Supplementary Material (SM)~\cite{sm}).
In the following, we assume periodic boundary conditions and connect
the SFF's ends to form a ring.
Therewith we avoid filament rotation like in vortices~\cite{sumino2012}
which are not relevant within the cell context. 

{\em Cross-linkers} --
We consider two types of cross-linkers which are connected to some static background filaments and 
induce SFF shape fluctuations. The background filaments are all perpendicular to the SFF,
with a lattice spacing $d_{\rm{mesh}}$ (see Fig. \ref{fig:MTModel}).

{\it Thermal} cross-linkers 
(see Fig.~S1 in SM~\cite{sm})
are permanently bound to the SFF. They step in both directions along the background network and thereby 
alter the SFF's shape. A step of a cross-linker, and hence the new SFF shape, is
accepted according to the detailed balance condition, i.e., with probability
$\min (1, \exp (-\Delta E/ k_BT))$,
where $\Delta E$ is the associated energy change and $k_B$ the Boltzmann constant.

{\it Active} cross-linkers may bind to or unbind from the SFF. 
Their binding is not direct but via a small infinitely flexible chain with maximum length $l_{\rm{max}}$.
The linker can exert a force only when its chain is extended.
An unbound active linker attaches
to an available binding
site (i.e., an intersection point between the SFF and one background filament)
with constant rate $\omega_{\rm a}$.
For each attachment event,
the stepping direction $(+)$ or $(-)$ of the cross-linker is randomly chosen and kept 
fixed until it detaches again.

Once attached, the active cross-linker stochastically takes discrete steps
along the background filament in the direction determined above.
The stepping rate $p(F_{\rm{SFF}})$ depends strongly on the load force
$F_{\rm{SFF}}$ exerted by the SFF on the linker
(see formulas (\ref{eq:s1}-\ref{eq:s2})
in SM~\cite{sm}).
If the load force pulls in opposite direction to the stepping
direction of the active linker,
the linker velocity is reduced. It stops 
when the load force exceeds the linker's stall force $F_{\rm{s}}$.
The active linkers stochastically detach with rate
$\omega_{\rm d}(F_{\rm{SFF}})= \omega_{{\rm d}_0} \exp\left(\frac{|F_{\rm{SFF}}|}{F_{\rm{d}}}\right)$
where $F_{\rm{d}}$ gives the detachment force scale.

Bound cross-linkers with an extended chain can exert a force
and possibly deform the SFF, which in turn will apply a restoring
force on the linkers. For some deformations of the SFF the restoring force may induce sudden 
detachments of cross-linkers or even initiate detachment cascades. 

The time scale separation of the linker dynamics and
SFF relaxation allows the simulation of single linker
activity and consecutive instantaneous SFF relaxation.
The coupling between the dynamics of SFF and cross-linkers
is implemented as follows for both types of linkers
(see SM, Section IV~\cite{sm} for more details).

{\em Equilibrium shape of a constrained SFF} --
The semi-flexible filament's shape is chosen to minimize the bending energy
under the constraints imposed by the pulling cross-linkers.
Between two 
consecutive pulling cross-linkers located in $x_i$ and $x_{i+1}$ the SFF shape is given by a profile $u_i(x)$, that minimizes the energy
(\ref{eq:defE})
of this portion of the SFF 
\begin{equation}
E_i=k \int_{x_i}^{x_{i+1}} \left[\partial_x^2 u_i(x)\right]^2dx \label{eq:E_WLC}
\end{equation}
assuming no overhang and $\left|\partial_x u_i(x) \right| \ll1$.

The force per unit length is $F\sim \partial^4_x u_i(x)$,
which 
vanishes
at equilibrium between two attachment points.
Thus the equilibrium is given by (see for instance Ref.~\cite{schwarz_m2001})
\begin{equation}
u_i(x)=a_i(x-x_i)^3+b_i(x-x_i)^2+c_i(x-x_i)+d_i
\end{equation}
for $ x_i\le x\le x_{i+1}$.
Let $z_i$ be the vertical displacement of the SFF imposed at position $x_i$ and  $v_i$ the local slope. 
The global SFF shape is given by the sequence of  single segments respecting the boundary constrains 
to ensure the differentiability of the global polynomial:
\begin{eqnarray}
u_i(x_i)=z_i, &\quad& u_i(x_{i+1})=z_{i+1}\\
\partial_x u_i(x_i)=v_i,&\quad& \partial_x u_i(x_{i+1})=v_{i+1}\; .
\label{eq:bc}
\end{eqnarray}
As detailed in SM, Section III~\cite{sm}, the coefficients
of the polynomial
can be expressed in terms of the constraints in $x_i$ and $x_{i+1}$.

Now if we consider that several of these segments are
put end-to-end, we have to minimize the global energy.
This minimization will determine the slopes at attachment points - and
thus the whole shape.
The detailed calculation is given in SM, Section III~\cite{sm}.
One important point is that, as a byproduct of the calculation,
we obtain also the local forces exerted on the pulling cross-linkers.

The calculation above requires that the vertical displacements $z_i$
at positions $x_i$ are known.
When a cross-linker is pulling, its chain is elongated.
Thus, if $h_i$ is the vertical position of the cross-linker,
we have $z_i = h_i \pm l_{\rm{max}}$, the sign depending on the direction
of the pulling force.
However, the semi-flexibility causes non-trivial response in terms of
SFF geometry and forces.
As the action of a single linker may change the SFF's global shape, we
need to check after
each relaxation of the SFF shape whether there is a change in the number
of pulling cross-linkers, and in that case reevaluate the SFF shape that
minimizes the bending energy.
An iterative procedure alternatively adjusting the coupling chains and relaxing the SFF
shape allows to converge towards the full equilibrium of the system.
Eventually, a procedure described in SM, Section V~\cite{sm} allows to keep the
length of the non tensile SFF constant.

{\em Avalanches} --
A characteristics of SFFs 
is that small changes in the applied forces 
may lead to large deformations of the SFF.
For our model this means that a 
single motor step
may change considerably the shape of the SFF,
and therefore the restoring 
forces, which may become so high that a subset of motors
will instantaneously detach.
Such a
detachment avalanche is done iteratively beginning with the linker that bears
the largest restoring force. The SFF shape and the forces are re-estimated after
each detachment event.

% ---------------- RESULTS ---------------- RESULTS ---------------- RESULTS ---------------- RESULTS ---------------- RESULTS ---------------- RESULTS ---------------- RESULTS ---------------- RESULTS ---------------- %
%
Now that our model is defined, we shall present some
numerical results on the SFF's shape characteristics under
coupled SFF-linkers dynamics.

{\em Bending spectrum and persistence length} --
Following~\cite{gittes1993,brangwynne2007b},
we analyze the fluctuations of the SFF shape $\theta(s)$
by using a decomposition into cosine modes
$\theta(s)=\sqrt{2/L} \sum_{n=0}^\infty a(q) \cos(q s)$,
with the wavenumber $q=\left(n\pi/L\right)$.

For {\em purely thermal} fluctuations in 2D with Boltzmann weight
$\propto e^{[- E/(k_BT)]}$ and 
$E$ given by Eq.~(\ref{eq:defE}),
the variance of cosine modes' amplitudes is
known to vary with $q$ as
\begin{equation}
\text{Var}(a(q))  \equiv
\langle a(q)^2 \rangle
= \frac{1}{\Lp}\frac{1}{q^2} \;
\;\;\; \mbox{with} \;\;
\Lp^{thermal}=\frac{2k}{k_{\text{B}}T}\;.
\label{eq:lpt}
\end{equation}
For other types of fluctuations, if the bending spectrum
has also a full $q^{-2}$ dependence,
one can still extract the persistence length by a simple fit
as in (\ref{eq:lpt}).
However, for
a more general case, a more direct definition of the persistence length
is given from the two point correlation function of the
tangent angle~$\theta$ \cite{landau_l1981a}:
\begin{equation}
\langle \cos\left(\theta(s)-\theta(s')\right)\rangle = \exp\left(-|s-s'|/(2\Lp)\right)
\label{eq:lp}
\end{equation}
for two dimensional fluctuations
(see Section VI in SM~\cite{sm} for more details).

\begin{figure}[t!]
	\includegraphics[width=\columnwidth]{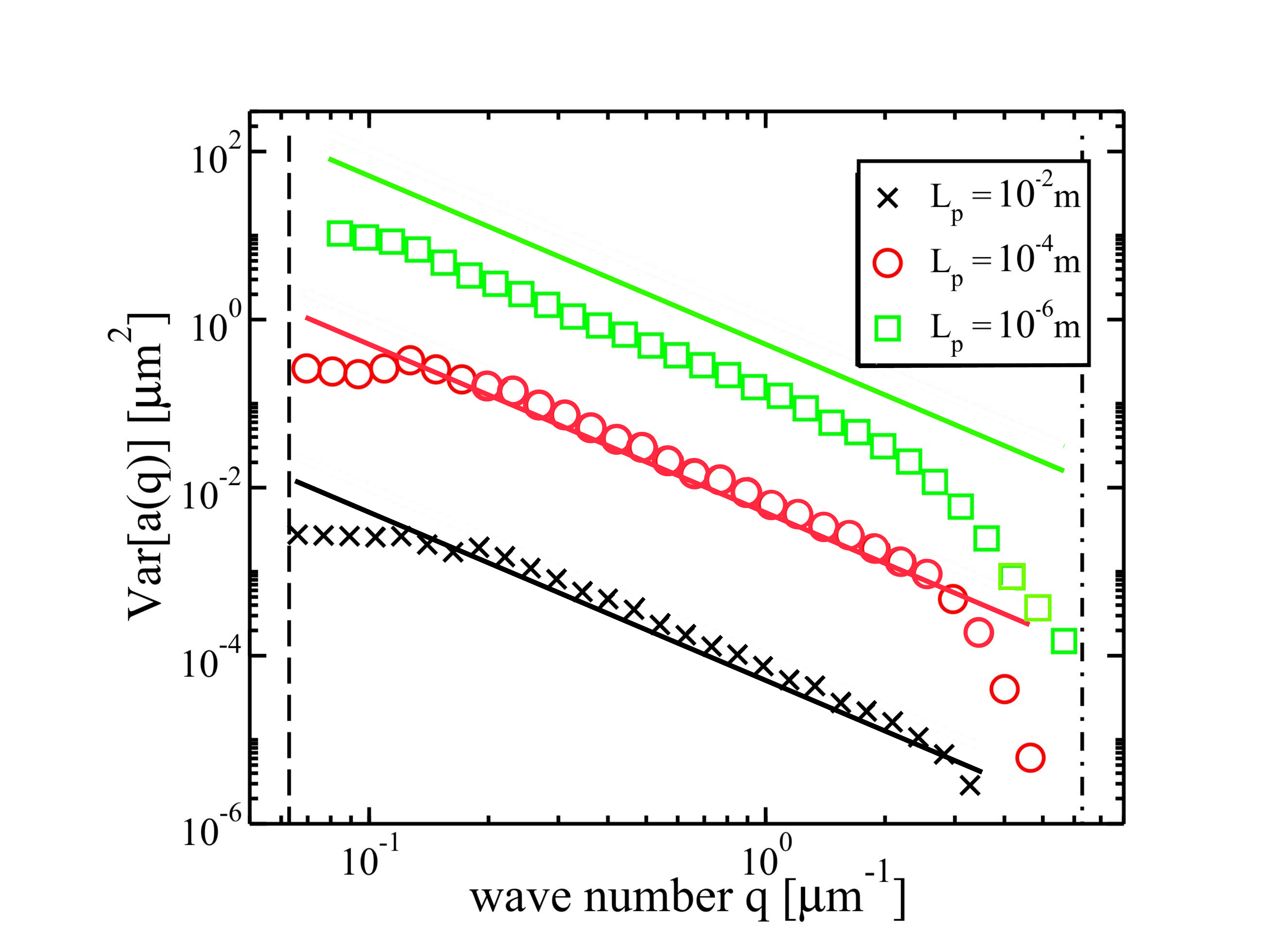}
	\caption{\label{pic:thermal} 
	Variance of the 
	amplitudes $a(q)$ for
	a SFF
	pulled by thermal linkers
	(symbols), for various bending rigidities
	(or, equivalently, various persistence lengths).
	The straight lines give the purely thermal bending spectra
	for the same persistence lengths (see Eq.~(\ref{eq:lpt})).
	}
\end{figure}

In our model, the fluctuations enforced through the active linkers
differ strongly from purely thermal fluctuations:
First, the fluctuations are transmitted not continuously in space but only at the binding sites of the
cross-linkers. Second, the dynamics of active cross-linkers does not fulfill detailed-balance, as discussed previously.
In order to disentangle these two effects,
we start our analysis with thermal cross-linkers rather than
with active ones.

\begin{figure}[t!]
	\includegraphics[width=\columnwidth]{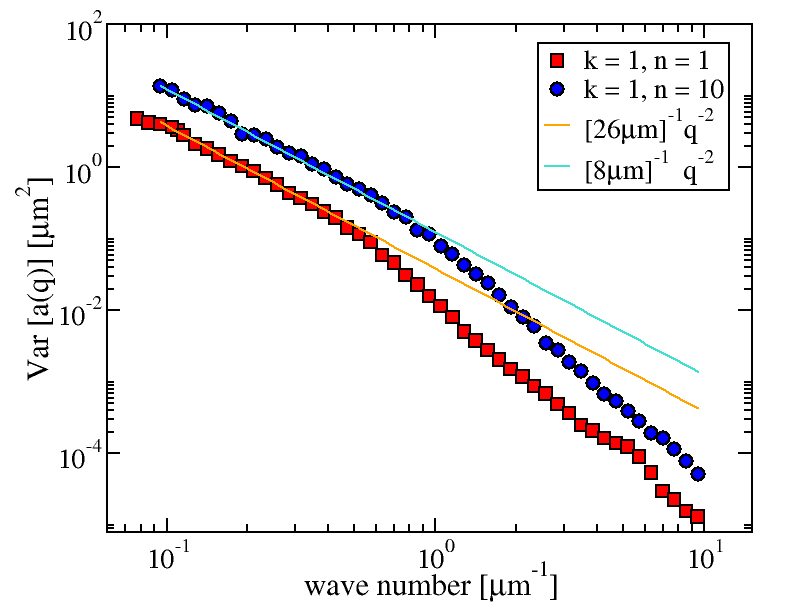}
\caption{\label{pic:driven_spectrum_k1} 
	Variance of the
	amplitudes $a(q)$ for
a SFF deformed by active linkers. For the red squares, the parameters are those of
Table 1 (SM~\cite{sm}).
The data follow a $q^{-2}$ behavior (orange line) for small $q$, which can be associated with a persistence length of $26\mu$m.
However we find a significant deviation from the $q^{-2}$ spectrum for larger
wave vectors.
An increase by a factor of $n=10$ of both $F_d$ and $F_s$
(blue circles,
stronger motors) extends the $q^{-2}$ regime, but does not suppress the
deviation completely.
}
\end{figure} 

{\em Thermal cross-linkers} --
In Fig.~\ref{pic:thermal} we 
show the cosine bending spectrum for thermal linkers, for three different bending rigidities and compare it to the purely thermal spectrum. 
For all bending rigidities we observe deviations from the purely thermal spectrum for small values of $q\sim 1/L$. These 
can be attributed to the finite length of the SFF. Modes for large wavenumbers  $q\sim 1/d_{\rm mesh}$ are also 
suppressed due to the finite distance between neighboring cross-linkers. For intermediate values of $q$ one 
indeed observes the expected $q^{-2}$ spectrum.
As long as the persistence length is
at least of the order of the system size,
we are in the regime of small deviations and
one obtains the expected value of the persistence length.
In the 
context of biological applications it is interesting to notice that the mesh size of the background lattice and the 
typical length of the SFF determine the range of the  $q^{-2}$ spectrum. 
 
\begin{figure}[t!]
	\includegraphics[width=\columnwidth]{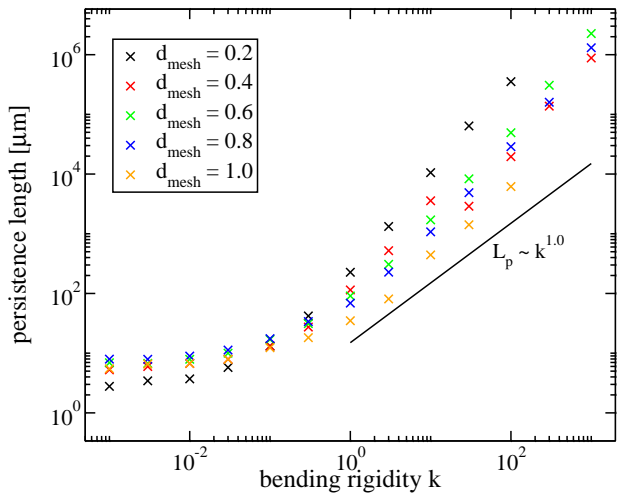}
\caption{\label{pic:Lp_meshsize_kVar} Persistence length as a function of the
bending rigidity for various background network mesh
size~$d_{\rm{mesh}}$ (in $\mu$m) and for active linkers.
For comparison, the straight line gives the linear increase expected for purely
thermal fluctuations (Eq. \ref{eq:lpt}).
}
\end{figure}

{\em Active cross-linkers} --
As a second step of our analysis, we now consider model with active cross-linkers. For active 
cross-linkers one observes generically strong deviations of the bending spectrum from $q^{-2}$ even in the 
regime $1/L < q < 1/d_{\rm mesh}$. The actual form of the spectrum depends strongly on the model 
parameters. In Fig.~\ref{pic:driven_spectrum_k1} we show bending
spectra for SFF with $k=1$, 
i.e. the bending rigidity of microtubules. For this value of $k$ and realistic biological parameters for 
the linker forces and mesh size, we find a thermal like $q^{-2}$ spectrum for small wave vectors 
while deviations arise for larger wave numbers. The extension of the $q^{-2}$ regime is even larger if  
$F_d$ and $F_s$ are scaled up by a factor of $n=10$, i.e. if linkers are made stronger. Using the $q^{-2}$ part of the 
bending
spectrum we obtain an apparent persistence length of $26\mu m$, surprisingly close to the 
experimentally obtained value of $30\mu m$ for \emph{in vivo} microtubule fluctuations \cite{brangwynne2007b}. 

As mentioned above, generically we observe rather strong deviations from the $q^{-2}$ regime.
Therefore, from now on, the persistence is estimated via the tangent angle
correlation (\ref{eq:lp})  to ensure reliable persistence length estimation for all
bending rigidities. 
In the case of purely thermal fluctuations, the persistence length is
proportional to the bending rigidity.
Fig.~\ref{pic:Lp_meshsize_kVar} reveals that
the active linker-driven SFF's persistent length evolves in a more complex way.
For small~$k$ we observe only a weak dependency of the apparent SFF
stiffness on the bending rigidity, as the deformations are limited by the mesh size.
An increase of the bending rigidity to $k\ge 1$ leads to a super-linear
increase of the persistence length up to, and beyond the SFF length.

Finally we also studied the effect of varying the mesh size of
the underlying network, for various bending rigidities of the SFF
and a fixed number of active-linkers.
As seen in Fig.~\ref{pic:Lp_meshsize_kVar},
for low bending rigidities, the persistence length slightly
decreases with $d_{\rm{mesh}}$
(This dependence is linear in $d_{\rm{mesh}}$, as can be seen
in Fig.~S2 in SM~\cite{sm}).
Though we use here the model beyond the limit of small
deformations, we expect this conclusion to hold.
Indeed, closer linkers can enforce deformations at smaller scales.

Surprisingly this behavior 
is inverted for large bending rigidities,
where one observes larger persistence lengths 
for higher densities of active linkers.
Indeed, when $d_{\rm{mesh}}$ decreases,
the curvature induced by a single linker step
is more pronounced.
At high bending rigidities, this strong local
deformation will result into strong load forces,
which most likely the linker will not be able
to sustain.
Therefore, it is difficult to deform the stiff
SFF at all,
if the density of cross-linkers
is too high.

% ---------------- DISCUSSION ---------------- DISCUSSION ---------------- DISCUSSION ---------------- DISCUSSION ---------------- DISCUSSION ---------------- DISCUSSION ---------------- DISCUSSION%

{\em Discussion}--
In this paper we have proposed a modeling approach to describe the dynamics of a semi-flexible filament subject
to fluctuations generated by a finite set of cross-linkers.
For any configuration of the linkers, we are able to compute the equilibrium shape of the semi-flexible filament,
using a semi-analytical method,
and also
to calculate the feedback on the linkers dynamics due to the SFF rigidity. This
allows us
to study quantitatively the effect of 
cross-linker induced deformations on the shape of the SFF
for
various types of linker
dynamics.

In this work we considered fluctuations generated by thermal and active linkers,
where the dynamics of the latter is based on the dynamics of typical kinesin motors.
In both cases linkers step perpendicular to the SFF. For thermal linkers we observe a $q^{-2}$ regime in the bending spectrum, whose range is limited by the length of the SFF for
small $q$ and by the distance between two linkers for large $q$.

In the case of active cross-linkers, one observes typically strong deviations from the  $q^{-2}$ bending spectrum, as
a signature of non-thermal fluctuations, in agreement with
experimental observations~\cite{brangwynne2008}. For biologically relevant parameters, however, a $q^{-2}$ spectrum has been
observed in a certain interval of wave-numbers. Interestingly, using this part of the bending spectrum one obtains
an estimate of the persistence length which is very close to the experimental value found in \cite{brangwynne2007b}. 
This agreement is remarkable, in view of the fact that we only assumed fluctuations perpendicular to the filament and
motor based microscopic dynamics.

Regarding the dynamics of the motor-driven SFF, our simulation results show the absence 
of bidirectional persistent displacement, which could be expected from a tug-of-war scenario~\cite{mueller_k_l2008}. 
These results are in agreement with explicit transport models~\cite{kunwar2011,klein_a_s2015a}.

Our approach can be generalized to other types
of non-thermal forcing and boundary conditions
and thus could be used in order to describe a large range of experimental settings.

%%%%%%%%%%%%%%%%%%%%%%%%%%%%%%%%%%%%%%%%%%%%%%%%%%%%%%%%%%%%%%%%%%%%%%%%%%%%%%
 %

\begin{acknowledgements}
We acknowledge CNRS support for a one month stay of L.S. at LPT. L.S. acknowledges 
support from the 
This work was supported by the Deutsche Forschungsgemeinschaft (DFG) within the 
collaborative research center SFB 1027 and the research training group GRK 1276. 
\end{acknowledgements}
%\input acknowledgement.tex   % input acknowledgement

%\bibliography{motors,neq,trafic,ped,statphys,bio,animals,fungi,livres,elast}
% none of the lines below should be decommented, else it does not work
%\bibliographystyle{apsrev4-1}
%\bibliographystyle{apsrev4}

\pagebreak
%%%%%%%%%% Merge with supplemental materials %%%%%%%%%%
%%%%%%%%%% Prefix a "S" to all equations, figures, tables and reset the counter %%%%%%%%%%
\setcounter{equation}{0}
\setcounter{figure}{0}
\setcounter{table}{0}
\setcounter{page}{1}
\makeatletter
\renewcommand{\theequation}{S\arabic{equation}}
\renewcommand{\thefigure}{S\arabic{figure}}
\renewcommand{\thesection}{\Roman{section}}

\widetext
\begin{center}
\textbf{\large Supplementary Material\\Non-equilibrium fluctuations of a semi-flexible filament
 driven by active cross-linkers}
\end{center}

\begin{equation}
p(F_{\rm{SFF}}) = \rm{max}\left[0,p_0\left (1- \frac{|F_{\rm{SFF}}|}{F_{\rm{s}}}\right)\right].
\label{eq:s1}
\end{equation}
On the contrary, stepping rate increases for a load force in the same
direction as the stepping
\begin{equation}
p(F_{\rm{SFF}}) = \rm{min }\left[2p_0,p_0\left (1+ \frac{|F_{\rm{SFF}}|}{F_{\rm{s}}}\right)\right].
\label{eq:s2}
\end{equation}

Though our model is an idealized one, we still used some
parameters corresponding to biological measurements,
as given in the table below.
In particular, the linkers' parameters are close to the kinesin ones.
The reference bending rigidity is the experimentally
measured value $k = 1.10^{-23}$Nm$^2$ for microtubules \cite{gittes1993}.

\bigskip

\begin{table}[h!]
\begin{minipage}{0.8\linewidth}
\begin{ruledtabular}
\begin{tabular}{lll}
\textit{Semi-flexible filament parameters}	&  & \textit{Value}\\
\hline
Bending rigidity in $k_{MT}$ units & $k$     			& $1 \, k_{MT}$\\
Bending rigidity of microtubules  & $k_{MT}$     			& $1.10^{-23} \, \text{Nm}^2$\\
Background mesh size & $d_{\rm{mesh}}$			& $1$ $\mu$m \\
Filament contour length 	& $L$ 			& $100$ $\mu$m \\
\hline
\hline
\textit{Cross linkers parameters }	&  & \textit{Value}\\
\hline
Binding rate 		& $\omega_{\rm{a}}$ 		&  $0.05 \, \text{s}^{-1}$\\
Unbinding rate 		& $\omega_{\rm{d}_0}$ 	&  $0.01 \, \text{s}^{-1}$ \\
Hopping rate 		& $p_0$  			&  $1 \, \text{s}^{-1}$ \\
Detachment force	& $F_{\rm{d}}$  		& 3 pN  \\
Stall force			& $F_{\rm{s}}$   	& 6 pN\\
Step size  			& $\delta$   	& 10 nm\\
Length of coupling chain & $l_{\rm{max}}$ & 10 steps\vspace{0.2cm}\\
\end{tabular}
\end{ruledtabular}
\end{minipage}
\caption{\label{tab:rates}System parameters and single active cross-linker characteristics.}
\end{table}

%\bigskip
%{\bf Polynomial description of the semi flexible filament shape: Bending energy and local forces}\\
\section{Polynomial description of the semi flexible filament shape: Bending energy and local forces}
\label{S3}

As stated in the core of the paper, the equilibrium shape of the SFF
between two attachment points can be taken under the form of a polymer of degree 3 (see Eq. (3) of the main text),
with the boundary conditions given by Eqs. (4) and (5) of the main text.
Putting all these polynomials end-to-end to construct the whole SFF,
this allows us to determine the polynomial's coefficients
 \begin{eqnarray}
a_i  &=& -2 \frac{\Delta z_i}{\Delta x_i^3} +  \frac{1}{\Delta x_i^2} \left( 2 v_i+\Delta v_i\right)\\
b_i  &=& 3 \frac{\Delta z_i}{\Delta x_i^2} -  \frac{1}{\Delta x_i} \left( 3 v_i+\Delta v_i\right)\\
c_i &=& v_i\\
d_i &=& z_i.
\end{eqnarray}
Here $\Delta x_i=x_{i+1}-x_{i}$ is the distance of two consecutive pulling cross-linkers, $\Delta z_i=z_{i+1}-z_{i}$  their vertical displacement difference and $\Delta v_i=v_{i+1}-v_{i}$ the difference of slopes. For simplicity, we admit periodic boundary conditions.
The energy of the SFF 
\begin{equation}\label{energy_quadratic}
E=z^t \tilde Bz-v^t\Lambda+v^t\, \tilde A \,v \;,
\end{equation}
where the matrices $\tilde A$ and  $\tilde B$ are given by
 \begin{eqnarray}
 \tilde{A}_{ij}&=&4 k \left( \frac{1}{\Delta x_i}+\frac{1}{\Delta x_{i-1}}\right),  \quad  \text{if  } i=j\nonumber\\
 &=&  \frac{2 k }{x_{\max (i,j)}-x_{\min (i,j)}}, \quad\text{if  } i=j\pm1\nonumber\\
&=& 0 \quad \text{else} 
  \end{eqnarray}
and
 \begin{eqnarray}
 \tilde{B}_{ij}&=& 12 k \left( \frac{1}{\Delta x_i^3}+\frac{1}{\Delta x_{i-1}^3}\right), \quad  \text{if }i=j,\nonumber\\
&=& -\frac{12 k }{(x_{\max (i,j)}-x_{\min (i,j)})^3}, \quad \text{if } i=j\pm1\nonumber\\
&=&  0 \quad \text{else.} 
  \end{eqnarray}
while the components of the vector $\Lambda$ in Eq. (\ref{energy_quadratic}) are given by
\begin{eqnarray}
 \Lambda_i = 12\,k\, \left[\frac{\Delta z_i}{(\Delta x_i)^2} + \frac{\Delta z_{i-1}}{(\Delta x_{i-1})^2} \right] \;.
\end{eqnarray}
The gradient $v$ is chosen in such a way that it minimizes the total energy in (\ref{energy_quadratic}), which allows us to write 
\begin{equation}
\Lambda_i=2\sum_{j}\tilde{A}_{ji}v_j.
\end{equation}
The coupling matrix  for  slopes  $\tilde{A}$ and for the local displacements $\tilde{B}$ are both cyclic-tridiagonal and take the form
With the gradient $v_i=\frac{1}{2}\sum_{k}\tilde{A}_{ik}^{-1}\Lambda_k$ a simple form for the global SFF energy is
\begin{equation}
E = z^t\tilde{B}z - \frac{1}{4} \Lambda^t \tilde{A}^{-1} \Lambda
 \end{equation}
and the local force reads
%\greenw{Force as a function of E to be added}
%
 \begin{eqnarray}
F_k = \frac{\partial E}{\partial z_k} &=& 24 k \left(\frac{\Delta z_{k-1}}{\Delta x_{k-1}^3}-\frac{\Delta z_{k}}{\Delta x_{k}^3}\right)\nonumber\\
&&- 12 k \left\lbrace v_k\left(\frac{1}{\Delta x_{k-1}^2}-\frac{1}{\Delta x_{k}^2}\right)\right\rbrace\nonumber\\
&&- 12 k \left\lbrace \frac{v_{k-1}}{\Delta x_{k-1}^2}-\frac{v_{k+1}}{\Delta x_{k}^2} \right\rbrace.
 \end{eqnarray}

%\bigskip
%{\bf Update Algorithm}\\
\section{Update Algorithm}
\label{S4}

Let us assume we start in a state for which we know
the force applied on each bound cross-linker,
and thus all the transition rates
(a possible initial state can be a flat SFF with no cross-linker
attached).

\begin{itemize}
\item We update the system of linkers with a tower sampling algorithm
and perform stochastic events until the occurrence of an event
that modifies the force exerted on the SFF.
\item Then the new equilibrium shape of the SFF is calculated
as explained in the next section.
\item The new forces exerted on the linkers are obtained
and the value of force-dependent rates is calculated
for each linker.
\end{itemize}
This procedure is repeated to update the system.

%\bigskip
%{\bf No tensibility condition}\\
\section{No tensibility condition}
\label{S5}

Though we assume that deformations are small enough to have no overhang
in the SFF shape, still deformations can be large enough to change significantly
the length of the polymer.
As we assume a non tensile polymer, we have to rescale the SFF in order
to keep a constant length.
When the vertical deformation becomes important, the horizontal extension
of the SFF should decrease, and the other way round when the deformation
decreases. This is done by removing or adding an empty
attachment site at a randomly chosen position.
As this is a discrete adjustment, it cannot compensate completely
the length change, and a small rescaling in $x$ and $z$ direction
is needed to keep the SFF's contour length constant.

\section{Definition of the persistence length}
\label{S6}

A persistence length gives the typical distance
over which a polymer subject to a given type
of fluctuations is deformed.
The persistence length obviously depends on the fluctuations
which are applied.
When not stated, these fluctuations are in general
implicitly assumed to be {\em purely thermal} fluctuations.
However, a persistence length can {\em a priori}
be defined for any kind of fluctuations - including
the fluctuations due to cross-linkers as we consider
in this paper.

As stated in the main text (Eq.~(\ref{eq:lp})),
a general definition of persistence lengths
%which can be used for any type of fluctuations
can be
given from the two point correlation function of
the tangent angle $\theta(s)$.

However, it is important to realize that,
even when one considers purely thermal fluctuations,
it is important to distinguish two dimensional
and three dimensional fluctuations.
Indeed, the deformation will not be the same in these
two cases.
In the literature, one can find two strategies
to account for these differences.
\begin{itemize}
\item{(i)} either a unique definition is given in terms
of the two point correlation function of
the tangent angle $\theta(s)$, but
different values of the persistence length
will be found for purely thermal fluctuations
depending whether these fluctuations are applied
in a 2D or 3D space.
\item{(ii)} or a different definition is given depending
on the dimension of space, resulting in a unique
value of the persistence length for purely thermal
fluctuations.
\end{itemize}
While the first point of view was used for example
in~\cite{kierfeld2008},
the second one was considered in~\cite{gittes1993,brangwynne2007b}.

In our paper, we chose to take the point of view (ii).
Then the definition of the persistence length for 3D fluctuations is
\begin{equation}
%\langle \theta(s)-\theta(s')\rangle = \exp\left(-|s-s'|/(2\Lp)\right)
\langle \cos\left(\theta(s)-\theta(s')\right)\rangle = \exp\left(-|s-s'|/\Lp\right),
\end{equation}
while a factor $2$ has to be added in front of $\Lp$
when considering two dimensional fluctuations, as given in Eq.~(\ref{eq:lp})
of the main text.
%The advantage of this choice of a dimension dependent
%definition of the persistence length
%is that it is consistent with the other definition
%that can be given in the case of purely thermal fluctuations, based
%on the $q^{-2}$ dependence of the variance of cosine modes' amplitudes
%(see Eq.~(\ref{eq:lpt}) in main text).
Then the definition of Eq.~(\ref{eq:lp}) is consistent with
the one of Eq.~(\ref{eq:lpt}).

%
%\bigskip
%{\bf Deformations under active linkers}\\
\section{Deformations under active linkers}
\label{S7}

\begin{figure}[h!]
	\includegraphics[width=0.6\columnwidth]{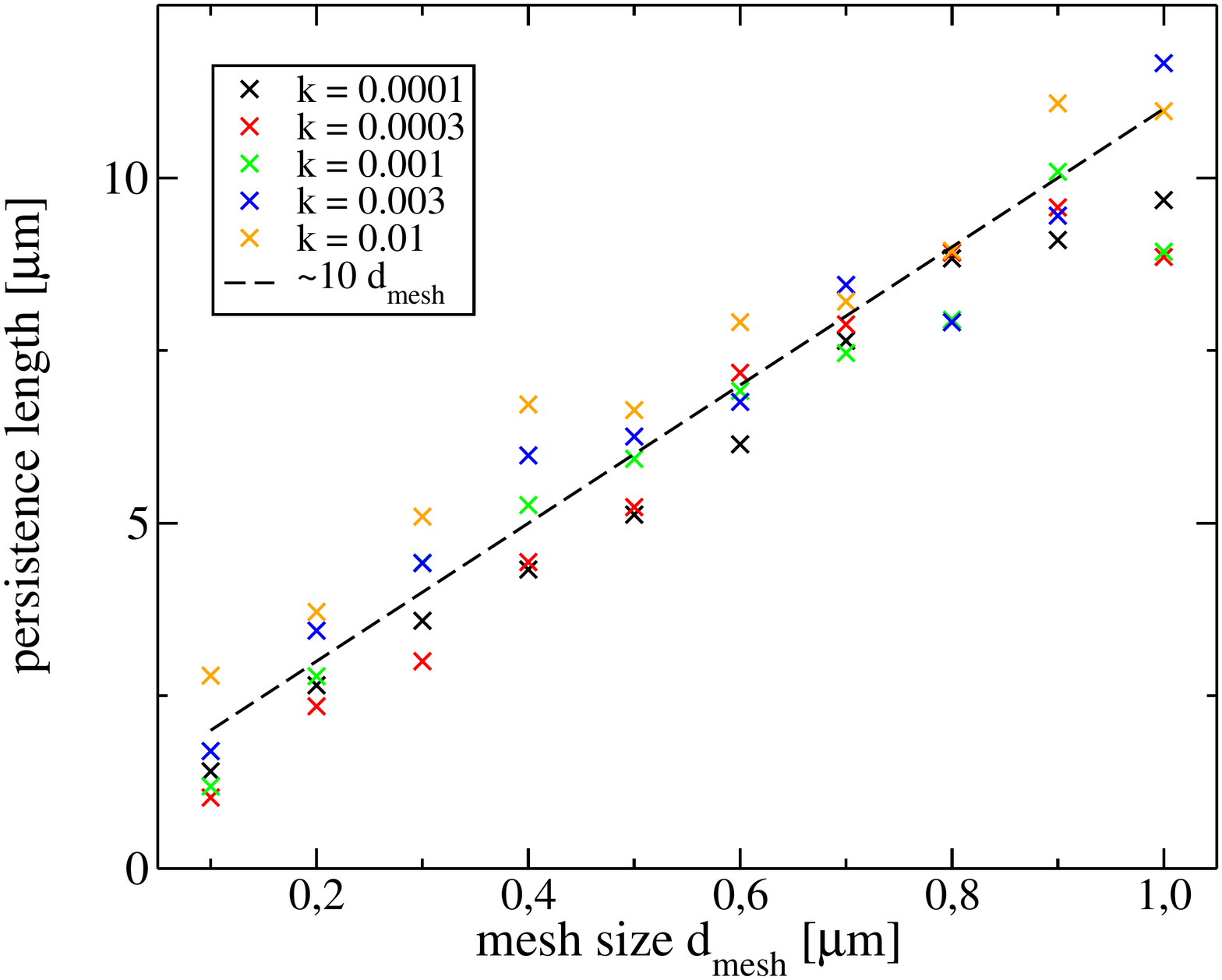}
	\caption{\label{pic:Lp_meshsize} Persistence length as a function of the background network mesh size~$d_{\rm{mesh}}$ in the case of active linkers. For flexible SFFs we observe a linear increase of the apparent stiffness with increasing mesh size.}
\end{figure}

In Fig.~\ref{pic:Lp_meshsize}, we show the linear dependence of the
persistence length as a function of the background network mesh size~$d_{\rm{mesh}}$.

\end{document}